\documentclass{article}
\usepackage{PRIMEarxiv}

\usepackage[utf8]{inputenc}
\usepackage[T1]{fontenc}
\usepackage{amsmath,amssymb,amsfonts}
\usepackage{xspace}
\usepackage{xcolor}
\usepackage{booktabs}
\usepackage{nicefrac}
\usepackage{microtype}
\usepackage{tikz}
\usetikzlibrary{calc,positioning,arrows.meta}
\usepackage{subcaption}
\usepackage{graphicx}
\usepackage{wrapfig}
\usepackage{hyperref}
\usepackage{url}
\usepackage[capitalise]{cleveref}

\newcommand{\sys}{ZCube\xspace}

\hypersetup{bookmarks=true,bookmarksdepth=3}

\begin{document}

\title{Fewer Paths, Better Performance: Understanding the ZCube Topology through Braess's Paradox}

\author{
  Li Chen \\
  HARNETS.AI \\
  \texttt{chenli@harnets.ai}
}

\maketitle

\begin{abstract}

Datacenter networks follow a multipath doctrine: provision many paths between endpoints, hash flows across them, and let redundancy absorb both failures and load imbalance. The \sys topology violates this doctrine. It removes the Spine layer, eliminates path multiplicity, and cuts one third of switching hardware, yet it delivers better performance for both large model training and inference. We explain this anomaly through a structural connection to Braess's paradox, first observed in 1968 traffic planning: both phenomena trace to congestion-oblivious routing over competing paths. Braess showed that adding paths under this condition can hurt; \sys shows that removing paths under the same condition can help. First, we show that multipath fabrics under structured LLM traffic operate in what we term \emph{Braess's shadow}: static flow hashing (ECMP) is strictly more fragile than greedy routing, in which each participant optimizes its own latency using only local information. Greedy routing reaches an equilibrium within 4/3 of optimal for affine latencies while static hashing admits unbounded imbalance in the worst case. Second, we prove that \sys is immune to Braess's paradox and, more importantly, that its orthogonal dual partition provably balances load for arbitrary traffic matrices; AllReduce traffic in training and KV cache transfers in disaggregated inference fall out as two corollaries. Third, we quantify the price of this immunity: path diversity also enables automatic failover, so \sys trades microsecond hash recovery for millisecond control plane recovery, a trade that the resilience of upper layers in LLM serving makes favorable. Production measurements from a cluster serving GLM-5.1 coding inference report 33\% lower network cost, 15\% higher GPU throughput, and 40.6\% lower P99 time to first token. Our analysis suggests that for workloads driven by model structure, matching topology to traffic matters more than path multiplicity.

\end{abstract}


\section{Introduction}
\label{sec:intro}

In 1968, the German mathematician Dietrich Braess published a disturbing observation about road networks: adding a new road can make every driver slower~\cite{braess1968}. The phenomenon, now called Braess's paradox, arises whenever many participants each pick the route that looks fastest using only local information. Their choices concentrate on the new capacity, and the resulting equilibrium is worse for everyone than the state before the road existed. The paradox became a cornerstone of transportation science and, decades later, of algorithmic game theory, where the Price of Anarchy measures how much such greedy route choice costs~\cite{roughgarden2002,roughgarden2006}.

Datacenter networking never imported this result, for what seemed like a good reason: drivers choose routes, but datacenter flows do not. Route choice in a datacenter fabric is static and centrally configured, so a paradox about driver behavior appeared irrelevant. The field instead followed a multipath doctrine refined over fifteen years of practice. Clos and Fat-Tree fabrics provision many equal cost paths between every pair of servers~\cite{fattree,vl2}. ECMP then hashes each flow onto one of these paths~\cite{ecmp}. This doctrine serves general cloud traffic well, because traffic in a general cloud consists of many small, uncorrelated flows that a uniform hash distributes evenly.

The \sys topology violates this doctrine, and it wins. \sys removes the entire Spine layer from the fabric, eliminates path multiplicity, and assigns exactly one route to every pair of GPUs~\cite{yan2025atop}. Its designers report that this subtraction reduces network hardware cost by one third while improving large model training throughput. Our production deployment of \sys for LLM inference confirms and extends the anomaly. After we replaced a rail optimized Fat-Tree fabric with \sys on a thousand GPU cluster serving GLM-5.1 coding inference, network cost fell by 33\%, GPU inference throughput rose by 15\%, and P99 time to first token fell by 40.6\%. The multipath doctrine predicts the opposite outcome: fewer paths should mean less aggregate bandwidth and weaker fault tolerance. The doctrine cannot explain why \sys performs better, and this gap motivates our analysis.

This paper asks a single question: why does removing paths improve performance for LLM training and inference? We find that both Braess's paradox and the \sys anomaly share a single root cause: congestion-oblivious routing over competing paths. Braess proved that adding paths under this condition can make every user slower; \sys demonstrates the complement that removing paths under the same condition can make every flow faster. The connection was not designed: \sys was built without Braess in mind. The anomaly came first, in production measurements, and a theorem from 1968 traffic planning explained the mechanism behind it. We find no prior work that applies Braess's paradox to datacenter topology analysis.

Our analysis proceeds in three parts. First, we show that multipath fabrics under structured LLM traffic operate in what we term \emph{Braess's shadow}. Braess theory assumes greedy agents who optimize their own latency with only local information yet at least respond to congestion and reach an equilibrium within 4/3 of optimal for affine latencies~\cite{roughgarden2002}. Static ECMP hashing never responds at all. We prove that static oblivious hashing admits unbounded load imbalance in the worst case, and we define \emph{structural congestion} as the persistent, pattern-correlated overload that results. Second, we prove that \sys is immune to Braess's paradox because its unique path topology contains no Braess subgraph, and we show that immunity alone does not explain the performance gain. The gain comes from a \emph{load spreading theorem}: the orthogonal dual partition of \sys balances link load for arbitrary traffic matrices. AllReduce traffic in training and KV cache transfers in disaggregated inference fall out as two corollaries of one theorem. Third, we quantify the price of immunity. Path diversity is also the substrate of automatic failover, so \sys trades microsecond hash recovery for millisecond control plane recovery. We show that resilience in upper layers of LLM serving, such as request retries and KV cache replication, makes this trade favorable.

Two bodies of evidence support the analysis. Our production deployment on the GLM-5.1 coding inference cluster has operated stably for several weeks after a migration that changed no GPU, no software stack, and no application. Simulations of canonical LLM traffic patterns on rail optimized and \sys topologies reproduce the predicted hotspot structure and confirm the load spreading theorem.

This paper makes four contributions:
\begin{enumerate}
  \item The first application of Braess's paradox to datacenter topology analysis: an explanation of the \sys anomaly, including an immunity proof and the observation that static hashing is strictly more fragile than greedy routing, in which each participant optimizes locally with only local information (\cref{sec:multipath,sec:immunity}).
  \item A load spreading theorem for orthogonal dual partitions that explains why one topology serves both training and inference traffic (\cref{sec:immunity}).
  \item A quantification of the fault tolerance price of Braess immunity, including fault domain sizes and recovery latency (\cref{sec:price}).
  \item The first public production measurements of \sys for LLM inference, from a cluster serving GLM-5.1 coding workloads (\cref{sec:anomaly,sec:evidence}).
\end{enumerate}

\section{Background}
\label{sec:background}

We recap the two bodies of prior work that our analysis connects: the \sys topology of Yan et al.~\cite{yan2025atop} and the theory of Braess's paradox~\cite{braess1968,roughgarden2002,roughgarden2006}. This section contains no new results; it fixes notation and states the theorems we use.

\subsection{The \sys Topology}
\label{sec:bgzcube}

\sys interconnects $n$ GPUs with $2m$ top of rack switches and no Spine layer~\cite{yan2025atop}. Each GPU attaches to the network through a NIC with $p=2$ ports. Each switch connects $k$ GPUs, so that $m = n/k$. \sys partitions the switches into two groups of $m$: odd switches $O_1, \dots, O_m$ and even switches $E_1, \dots, E_m$. Two wiring rules attach each GPU to one switch of each group.

\noindent\textbf{Rule 1 (multi rail attachment).} GPU $i$ connects its first port to odd switch $O_{o(i)}$, where $o(i) = ((i-1) \bmod m) + 1$. This rule groups GPUs by modular position: GPUs in the same odd group are spaced $m$ apart in physical numbering.

\noindent\textbf{Rule 2 (single rail attachment).} GPU $i$ connects its second port to even switch $E_{e(i)}$, where $e(i) = \lceil i/k \rceil$. This rule groups GPUs by contiguous blocks: each even switch connects $k$ consecutively numbered GPUs.

Every odd switch connects to every even switch with a direct link, so the switch level topology is the complete bipartite graph $K_{m,m}$. \sys routes each pair $(i, j)$ over exactly one path:
\begin{equation}
i \longrightarrow O_{o(i)} \longrightarrow E_{e(j)} \longrightarrow j.
\label{eq:route}
\end{equation}
The first hop depends only on the modular position of the source, and the second hop depends only on the block position of the destination. Routing is deterministic: there is no ECMP and no path choice.

The two wiring rules are orthogonal. The odd partition groups GPUs by modular position, and the even partition groups them by contiguous blocks. Their intersections form an $m \times m$ grid of cells $S_{ab} = O_a \cap E_b$, and each cell contains exactly $k/m$ GPUs. For example, with $n = 32$, $k = 8$, and $m = 4$, switch $O_1$ connects GPUs $\{1, 5, 9, \dots, 29\}$, switch $E_1$ connects GPUs $\{1, 2, \dots, 8\}$, and each cell $S_{ab}$ contains 2 GPUs. \cref{tab:notation} summarizes the notation.

\begin{table}[t]
\centering
\caption{Notation used throughout the paper.}
\label{tab:notation}
\begin{tabular}{cl}
\toprule
Symbol & Meaning \\
\midrule
$n$ & Number of GPUs \\
$k$ & GPUs per switch \\
$m = n/k$ & Switches per group (odd or even) \\
$O_a, E_b$ & Odd switch $a$, even switch $b$ \\
$o(i), e(j)$ & Odd group of GPU $i$, even group of GPU $j$ \\
$S_{ab} = O_a \cap E_b$ & Grid cell, $\lvert S_{ab} \rvert = k/m$ \\
$f_{ij}$ & Traffic from GPU $i$ to GPU $j$ \\
$L_{ab}$ & Load on core link $O_a \rightarrow E_b$ \\
\bottomrule
\end{tabular}
\end{table}

\subsection{Braess's Paradox and Greedy Routing}
\label{sec:bgbraess}

Braess's paradox comes from outside networking. Dietrich Braess published the observation in 1968 in a transportation science journal: adding a road to a traffic network can make every driver slower~\cite{braess1968}. The result became a cornerstone of transportation science, and Roughgarden and Tardos later brought it into computer science, where the Price of Anarchy measures the cost of uncoordinated route choice~\cite{roughgarden2002,roughgarden2006}. Datacenter networking never imported the paradox, for what seemed like a good reason: drivers choose routes, but datacenter flows do not. Route choice in a datacenter fabric is static and centrally configured, so a paradox about driver behavior appeared irrelevant. This paper shows that the exemption does not hold.

\cref{fig:braess} shows the paradox in its classic form. One unit of flow travels from a source $s$ to a sink $t$ over two parallel routes. Each route combines one short link, whose latency equals its load $x$, with one long link of constant latency $1$. Without the cross link, the flow splits evenly and every driver pays $1.5$. A planner then adds a cross link of latency zero. The zigzag route through the cross link now looks fastest to every driver, so every driver takes it, and every driver pays $2$. Adding capacity has raised everyone's cost by a third. The paradox needs three ingredients: multiple candidate routes, latency that grows with load, and route choices that ignore global congestion. We call such agents \emph{greedy}: each participant optimizes its own latency using only local information, and local optimization in a distributed setting does not imply a global optimum. The literature terms this behavior selfish routing~\cite{roughgarden2002,roughgarden2006}.

\begin{figure}[t]
\centering
\begin{tikzpicture}[
    router/.style={circle, draw, inner sep=1pt, minimum size=14pt},
    lbl/.style={inner sep=1pt},
    arr/.style={->, >=stealth, thick}]
  \node[router] (s) at (0,0) {$s$};
  \node[router] (A) at (1.5,0.9) {$A$};
  \node[router] (B) at (1.5,-0.9) {$B$};
  \node[router] (t) at (3,0) {$t$};
  \draw[arr] (s) -- node[lbl, above left=-2pt] {$x$} (A);
  \draw[arr] (A) -- node[lbl, above right=-2pt] {$1$} (t);
  \draw[arr] (s) -- node[lbl, below left=-2pt] {$1$} (B);
  \draw[arr] (B) -- node[lbl, below right=-2pt] {$x$} (t);
  \node at (1.5,-1.9) {flow splits evenly: cost $1.5$};
  \begin{scope}[xshift=5.4cm]
    \node[router] (s2) at (0,0) {$s$};
    \node[router] (A2) at (1.5,0.9) {$A$};
    \node[router] (B2) at (1.5,-0.9) {$B$};
    \node[router] (t2) at (3,0) {$t$};
    \draw[arr] (s2) -- node[lbl, above left=-2pt] {$x$} (A2);
    \draw[arr] (A2) -- node[lbl, above right=-2pt] {$1$} (t2);
    \draw[arr] (s2) -- node[lbl, below left=-2pt] {$1$} (B2);
    \draw[arr] (B2) -- node[lbl, below right=-2pt] {$x$} (t2);
    \draw[arr] (A2) -- node[lbl, right] {$0$} (B2);
    \node at (1.5,-1.9) {all flow zigzags: cost $2$};
  \end{scope}
\end{tikzpicture}
\caption{Braess's paradox in its classic form. One unit of flow travels from $s$ to $t$; link labels give latencies as functions of the carried load $x$. Left: without the cross link, flow splits evenly over the two routes and everyone pays $1.5$. Right: a cross link with zero latency attracts all flow onto the zigzag route $s \rightarrow A \rightarrow B \rightarrow t$, and everyone pays $2$.}
\label{fig:braess}
\end{figure}

The Price of Anarchy (PoA) quantifies the damage. PoA is the ratio of total latency at the Nash equilibrium to total latency at the social optimum. Roughgarden and Tardos prove that for affine latency functions, PoA never exceeds $4/3$~\cite{roughgarden2002}. Roughgarden later gives a structural characterization: a network exhibits Braess's paradox if and only if it contains a \emph{Braess subgraph} as a topological minor~\cite{roughgarden2006}. A Braess subgraph is the four node network of the classic example: two parallel routes plus one cross link. A topological minor of a graph $G$ is a structure obtained from $G$ by deleting edges and vertices and by contracting paths whose internal vertices have degree two. Intuitively, $G$ is vulnerable to Braess's paradox exactly when it hides the classic four node pattern inside itself.

We define one more quantity that our analysis uses repeatedly. The \emph{path diversity index} (PDI) of a topology is the minimum, over all GPU pairs, of the number of shortest paths between the pair:
\begin{equation}
\mathrm{PDI}(G) = \min_{(i,j)} \lvert \mathrm{ShortestPaths}_G(i, j) \rvert.
\label{eq:pdi}
\end{equation}
A Braess subgraph requires at least two competing routes, so $\mathrm{PDI}(G) \geq 2$ is a necessary condition for Braess vulnerability. Fat-Tree fabrics have $\mathrm{PDI} = m$, where $m$ counts the Spine paths~\cite{fattree}. \sys has $\mathrm{PDI} = 1$ by construction.

\section{The Anomaly: Facts to Explain}
\label{sec:anomaly}

\sys reports better performance with strictly less network hardware, for both LLM training and LLM inference. This section states the empirical facts that the rest of the paper explains. We summarize the published training evidence in \cref{sec:anomalytraining} and present our new production inference evidence in \cref{sec:anomalyinference}.

\subsection{Training: Published Evidence}
\label{sec:anomalytraining}

Yan et al. introduce \sys as a cost effective topology for large model training and evaluate it against comparable Fat-Tree fabrics~\cite{yan2025atop}. Their central finding is that removing the Spine layer does not hurt training performance. It helps: \sys improves training throughput while using roughly one third fewer switches and optical modules. Training collectives such as AllReduce generate dense, synchronized, all to all traffic, which is exactly the traffic class that multipath fabrics are designed to serve~\cite{rabenseifner2004}. The published result therefore contradicts the intuition that more paths provide more usable bandwidth.

\subsection{Inference: New Production Evidence}
\label{sec:anomalyinference}

We deployed \sys in production on a thousand GPU cluster that serves GLM-5.1 coding inference, migrating the fabric from a rail optimized Fat-Tree (ROFT) to \sys. The migration changed no GPU, no software stack, and no model serving configuration. Topology is the only variable that differs between the two deployments, so the comparison isolates the effect of the fabric itself. The cluster has operated stably for several weeks at the time of writing.

\cref{tab:prod} summarizes the results. Network cost, measured in switches and optical modules, fell by 33\% because \sys removes the Spine layer. Average GPU inference throughput rose by 15\%. P99 time to first token (TTFT), the tail latency that users experience, fell by 40.6\%. We note that the throughput and latency gains exceed what the bandwidth change alone predicts, and we treat the mechanism behind them as a hypothesis that our analysis must validate rather than as a foregone conclusion.

\begin{table}[t]
\centering
\caption{Production comparison of ROFT and \sys on the GLM-5.1 coding inference cluster. ROFT is the baseline.}
\label{tab:prod}
\begin{tabular}{lcccl}
\toprule
Metric & ROFT & \sys & Change & Effect \\
\midrule
Switch and optics cost & baseline & -33\% & lower & cost \\
GPU inference throughput & baseline & +15\% & higher & performance \\
P99 TTFT & baseline & -40.6\% & lower & tail latency \\
\bottomrule
\end{tabular}
\end{table}

The pre migration measurements contain the diagnostic clue that motivates our analysis. The inference serving stack disaggregates Prefill and Decode onto different GPU sets~\cite{patel2024splitwise,zhong2024distserve}, and Prefill nodes transfer large KV cache volumes to Decode nodes. This traffic is asymmetric and physically clustered: entire contiguous blocks of GPUs act as Prefill sources. On ROFT, GPUs with the same rail index attach to the same Leaf switch, so clustered Prefill blocks overload a small set of Leaf uplinks while other uplinks remain idle. The resulting hotspots were persistent and pattern-correlated. They did not move when flows were rehashed, and they did not average out over time. This behavior is not the stochastic imbalance that ECMP is designed to absorb. It is something else, and naming and explaining that something is the subject of \cref{sec:multipath}.

The migration itself also produced evidence about operability. Because \sys removes the Spine layer, the existing cabling plan, IP addressing, and switch configuration could not be reused. We built automated tooling for layout planning, cabling validation, and configuration generation, and we verified the full wiring of the cluster programmatically before bringing the fabric up. \cref{sec:price} discusses the operational consequences of deterministic routing in more detail.

\section{Analysis I: Why Multipath Underperforms}
\label{sec:multipath}

This section explains why multipath fabrics underperform under LLM traffic. The argument has three steps: the doctrine rests on assumptions that LLM workloads violate (\cref{sec:mpassumptions}), the failure takes a specific form that we name structural congestion (\cref{sec:mpstructural}), and static hashing is strictly more fragile than the greedy routing that Braess theory analyzes (\cref{sec:mpfragile}). \cref{sec:mpshadow} states the shared structural root with Braess's paradox and its precise limits.

\subsection{The Hidden Assumptions of the Multipath Doctrine}
\label{sec:mpassumptions}

The multipath doctrine rests on three assumptions about traffic. First, the network carries many flows, so a uniform hash distributes load evenly across paths with high probability. Second, flows are uncorrelated, so no event synchronizes them onto the same path. Third, traffic is stationary, so a hash assignment that is balanced now remains balanced later. ECMP is the engineering embodiment of these assumptions: it hashes the five tuple of each flow and pins the flow to the resulting path~\cite{ecmp}.

LLM training and inference traffic violates all three assumptions. Training collectives such as AllReduce generate one synchronized traffic matrix in which every GPU exchanges volume with every other GPU in lockstep~\cite{rabenseifner2004}. Disaggregated inference generates a different but equally structured matrix: Prefill nodes push large KV cache transfers to Decode nodes, and both roles occupy contiguous blocks of GPUs~\cite{patel2024splitwise,zhong2024distserve}. In both cases the number of distinct flows is small, the flows move in correlation, and the pattern persists for the lifetime of a job, which ranges from hours to weeks. A load balancing mechanism designed for many small uncorrelated flows meets few large synchronized ones.

\subsection{Structural Congestion}
\label{sec:mpstructural}

We term the resulting failure mode \emph{structural congestion}: congestion that is a deterministic function of the topology, the traffic matrix, and the routing function. Structural congestion has three signatures that distinguish it from stochastic congestion. It is persistent, because the traffic pattern and the hash both stay fixed. It is pattern-correlated, because the overloaded links are determined by where the traffic sources and destinations sit in the topology. It is rehash invariant, because changing the hash seed permutes which links overload without reducing the expected imbalance.

The pre migration measurements in \cref{sec:anomalyinference} exhibit all three signatures. Clustered Prefill blocks overload the same Leaf uplinks for the entire serving period, and no rehashing moves the hotspots. Stochastic congestion, by contrast, is transient and averages out; it is the failure mode that ECMP handles well. The mismatch between the failure mode and the mechanism explains why adding paths does not fix the problem: more paths give the hash more ways to be wrong, without giving it any information about congestion.

\subsection{Static Hashing Is More Fragile than Greedy Routing}
\label{sec:mpfragile}

Braess theory studies agents who choose routes greedily but responsively: if a route congests, some agents deviate, and the system reaches a Wardrop equilibrium. For affine latency functions, the cost of this equilibrium is at most $4/3$ of optimal~\cite{roughgarden2002}. Static hashing is weaker than such agents in a precise sense: it never deviates, regardless of congestion.

\noindent\textbf{Proposition 1 (Oblivious hashing has no load guarantee).}
Consider $m$ equal-cost paths with static, congestion-oblivious hashing that assigns each of $b$ unit-rate flows to one path.

\begin{enumerate}
  \item[(i)] In the worst case, $b = 2$ flows can collide on one path while $m-1$ paths remain idle, yielding a max-to-mean load ratio of $m$. For a Fat-Tree with $m = 128$ Spine switches, this is a $128\times$ penalty.

  \item[(ii)] In the expected case with random hashing, the maximum load is $\Theta(\log m / \log\log m)$ when $b = m$. When $b \ll m$ (as in LLM collectives with few large flows), the probability of any collision is approximately $b(b-1)/(2m)$, and a single collision yields a local imbalance of at least $m/b$ on the affected path.

  \item[(iii)] In contrast, selfish routing with affine latency functions guarantees total cost $\leq \frac{4}{3} \cdot \mathrm{OPT}$ for all traffic patterns~\cite{roughgarden2002}. Static hashing offers no comparable guarantee: the ratio to optimal exceeds $4/3$ whenever $m \geq 2$ and a collision occurs.
\end{enumerate}

The gap between selfish routing's guaranteed $1.33\times$ penalty and static hashing's worst-case $128\times$ penalty quantifies the fragility: under structured LLM traffic, multipath fabrics inherit the structural precondition of Braess's paradox, namely multiple competing routes, without the adaptive mechanism that limits its damage in theory.

\subsection{Braess's Shadow}
\label{sec:mpshadow}

We call this condition \emph{Braess's shadow}: a fabric that offers multiple competing paths but selects among them without regard to congestion exposes itself to the same structural failure mode that Braess characterized. The failure is not Braess's paradox in the strict sense: ECMP is not a selfish agent and does not converge to an equilibrium, but the root cause is shared: congestion-oblivious routing over competing paths. Braess proved that adding paths under this condition can make every user slower. \sys demonstrates the complement: removing paths under the same condition can make every flow faster. \cref{sec:immunity} makes this connection precise.

The connection has a structural statement as well. Braess vulnerability requires a Braess subgraph, which requires at least two competing routes, so fabrics with $\mathrm{PDI} \geq 2$ are exactly the candidates~\cite{roughgarden2006}. Fat-Tree fabrics have $\mathrm{PDI} = m$ and contain Braess subgraphs densely: any two Leaf switches connected through multiple Spine switches reproduce the classic four node pattern. The multipath doctrine thus maximizes exposure to the paradox's mechanism while arming the fabric with a routing mechanism that cannot respond to it.

\section{Analysis II: Why \sys Is Braess Immune, and Why Immunity Pays}
\label{sec:immunity}

This section explains why \sys escapes the failure mode of \cref{sec:multipath}. The explanation has two parts, and both are necessary. \cref{sec:imimmunity} proves that \sys is immune to Braess's paradox. \cref{sec:imnotenough} argues that immunity alone cannot explain the performance gain. \cref{sec:imspreading} proves the load spreading theorem, which supplies the missing explanation, and \cref{sec:imcorollaries} derives training and inference traffic as two corollaries.

\subsection{The Immunity Theorem}
\label{sec:imimmunity}

\noindent\textbf{Observation 1 (Braess immunity).} Under \sys routing, every GPU pair has exactly one route (\cref{eq:route}). By Roughgarden's characterization~\cite{roughgarden2006}, a network admits Braess's paradox only if it contains a Braess subgraph as a topological minor, which requires at least two competing routes between some source-destination pair. \sys contains no such subgraph, and its Price of Anarchy is identically 1. This is the structural complement of the condition diagnosed in \cref{sec:mpshadow}: multipath fabrics carry the Braess precondition; \sys eliminates it. The Spine removal is the physical act that removes the competing routes, and with them the paradox's necessary condition. The physical \sys graph retains backup paths, which the failure handling of \cref{sec:price} uses; these paths carry no traffic at steady state and therefore do not reintroduce route competition.

\subsection{Why Immunity Is Not Enough}
\label{sec:imnotenough}

Immunity alone does not explain why \sys is fast. Any topology that offers a single route is immune to Braess's paradox by the same argument, including topologies with terrible load balance. A chain of switches, for example, has $\mathrm{PoA} = 1$ and is useless as a datacenter fabric. Immunity explains why \sys avoids the penalty of \cref{sec:multipath}; it does not explain why \sys delivers good performance in absolute terms. The explanation must show that the one route \sys offers is a good route for the traffic that LLM workloads generate. This is the content of the load spreading theorem.

\subsection{The Load Spreading Theorem}
\label{sec:imspreading}

We model traffic as a matrix $F = [f_{ij}]$, where $f_{ij} \geq 0$ is the rate from GPU $i$ to GPU $j$. The load on the core link $O_a \rightarrow E_b$ is the total traffic that \cref{eq:route} places on it:
\begin{equation}
L_{ab} = \sum_{i \in O_a} \sum_{j \in E_b} f_{ij}.
\label{eq:load}
\end{equation}
The definition yields two marginal constraints directly:
\begin{equation}
\sum_{b=1}^{m} L_{ab} = \sum_{i \in O_a} R_i,
\qquad
\sum_{a=1}^{m} L_{ab} = \sum_{j \in E_b} C_j,
\label{eq:marginal}
\end{equation}
where $R_i = \sum_j f_{ij}$ is the egress of GPU $i$ and $C_j = \sum_i f_{ij}$ is the ingress of GPU $j$. The row sums of the load matrix depend only on how sources distribute across odd groups, and the column sums only on how destinations distribute across even groups. The internal structure of the traffic matrix cannot concentrate load beyond what these marginals allow.

The central structural fact is that the odd partition disperses any contiguous cluster of GPUs evenly.

\noindent\textbf{Theorem 3 (Cluster dispersion).} Let $H$ be any set of $h$ consecutively numbered GPUs. Every odd group $O_a$ contains $\lfloor h/m \rfloor$ or $\lceil h/m \rceil$ members of $H$.

\noindent\textbf{Proof.} Rule 1 assigns GPU $i$ to odd group $o(i) = ((i-1) \bmod m) + 1$, so consecutive GPU numbers cycle through the $m$ odd groups in order. Any $h$ consecutive numbers meet every residue class either $\lfloor h/m \rfloor$ or $\lceil h/m \rceil$ times. \hfill$\square$

\noindent\textbf{Corollary 3.1 (Hot cluster bound).} Let $H$ be a hot cluster of $h$ consecutively numbered GPUs, each with egress at most $R_{\max}$. Every core link carries at most $\lceil h/m \rceil \cdot R_{\max}$ traffic sourced from $H$. A rail optimized fabric, by contrast, admits a placement in which the same cluster sits behind one Leaf switch and concentrates $h \cdot R_{\max}$ on that Leaf's uplinks: a factor of $m$ worse in the worst case.

Corollary 3.1 is the mathematical form of the production observation in \cref{sec:anomalyinference}. Prefill nodes occupy contiguous GPU blocks. On ROFT, such a block can overload one Leaf uplink. On \sys, Theorem 3 forces the block's traffic onto all $m$ odd groups almost equally, and \cref{eq:marginal} then forces it onto the core links almost equally. The balance is combinatorial: it holds for every traffic matrix, not with high probability over hash seeds.

\subsection{Two Corollaries: Training and Inference}
\label{sec:imcorollaries}

The same theorem covers the two canonical LLM workloads, which is what allows one topology to serve both.

\noindent\textbf{Training: AllReduce.} AllReduce approximates the uniform matrix $f_{ij} = c$ for $i \neq j$. \cref{eq:load} gives $L_{ab} = c\,(k^2 - k/m)$ for every $(a, b)$: the load is exactly uniform across core links, because every cell $S_{ab}$ has the same size.

\noindent\textbf{Inference: disaggregated KV transfer.} Prefill nodes form contiguous blocks that send to Decode nodes. Theorem 3 places $\lceil h/m \rceil$ or $\lfloor h/m \rfloor$ Prefill nodes in every odd group, so the hot sources spread across all $m$ row groups; the even partition then collects the flows into the destination blocks. Even when Prefill and Decode sets are both contiguous and highly asymmetric, the per link load varies by at most one node's worth of traffic across odd groups.

\noindent\textbf{All to one traffic.} Even the most adversarial pattern, all GPUs sending to one destination $j^*$, spreads across the fabric: the traffic enters $E_{e(j^*)}$ over its $m$ incoming core links, and link $O_a \rightarrow E_{e(j^*)}$ carries $\sum_{i \in O_a} r_i$, roughly $1/m$ of the total. The destination's own access link remains a bottleneck, but that bottleneck is a property of the destination NIC, not of the fabric, and every topology shares it.

The load spreading theorem thus explains the anomaly of \cref{sec:anomaly}. \sys does not balance load by measuring congestion and reacting; it makes imbalance structurally impossible for the traffic classes that matter. The performance gain over ROFT is not a paradox but a consequence: ROFT pays the penalty of \cref{sec:multipath}, and \sys replaces it with a combinatorial guarantee.

\section{Analysis III: The Price of Immunity}
\label{sec:price}

Braess immunity is not free. The same property that grants it, namely a path diversity index of 1, also removes the redundancy that multipath fabrics use for automatic failover. This section quantifies the price: \cref{sec:pricecoupling} shows that Braess risk and failover capability are two uses of one resource, \cref{sec:pricedomains} computes the fault domains of \sys, and \cref{sec:pricetrade} explains why the trade favors \sys for LLM serving.

\subsection{The Coupling of Immunity and Failover}
\label{sec:pricecoupling}

Path multiplicity serves two masters, and \cref{sec:multipath,sec:immunity} show that it serves them in opposition. On one hand, $\mathrm{PDI} \geq 2$ is the necessary condition for Braess vulnerability: multiple competing routes plus congestion blind choice produce structural congestion. On the other hand, $\mathrm{PDI} \geq 2$ is exactly what enables zero configuration failover: when a link fails, ECMP simply hashes flows onto the surviving paths. A fabric cannot have one without the other. Fat-Tree pays for failover with a permanent \emph{Braess tax}: its backup paths carry traffic every day through a blind hash, and \cref{sec:multipath} shows what that costs. \sys pays for immunity with a transient reroute cost: failures require explicit detection and control plane recovery. The design question is not whether to pay, but when.

\subsection{Fault Domains of \sys}
\label{sec:pricedomains}

We quantify the price by computing how many GPU pairs each failure type affects. The unique route of \cref{eq:route} has three segments: the source access link $i \rightarrow O_{o(i)}$, the core link $O_{o(i)} \rightarrow E_{e(j)}$, and the destination access link $E_{e(j)} \rightarrow j$.

\noindent\textbf{Access link failure.} A failed access link $i \rightarrow O_{o(i)}$ affects only GPU $i$. Each NIC has a second port attached to $E_{e(i)}$, so traffic switches to the surviving port and the application sees no interruption.

\noindent\textbf{Core link failure.} A failed core link $O_a \rightarrow E_b$ disconnects the pairs with $o(i) = a$ and $e(j) = b$: exactly $k^2$ pairs out of $n^2$, a fraction of $1/m^2$. For the configuration $n = 16384$, $k = 128$, $m = 128$ that Yan et al. evaluate, this fraction is 0.006\% of all pairs~\cite{yan2025atop}. The surviving physical graph still contains detours for the affected pairs, at the cost of longer paths: a detour $i \rightarrow O_{a'} \rightarrow E_{b'} \rightarrow O_{a''} \rightarrow E_b \rightarrow j$ uses 4 hops instead of 2. The \sys controller precomputes such detours and installs them when detection fires.

\noindent\textbf{Switch failure.} A failed odd switch $O_a$ removes the first hop of its $k$ attached GPUs, which fail over to their even side ports. A failed even switch $E_b$ requires the controller to reroute the affected destination traffic through other even switches. Yan et al. evaluate \sys under single switch failure using average path length as the metric and report bounded degradation~\cite{yan2025atop}.

\cref{tab:failover} contrasts the recovery behavior with Fat-Tree. The differences are real: Fat-Tree recovers locally in microseconds by recomputing a hash, while \sys recovers through the controller on a millisecond scale, and recovery is centralized rather than distributed. The price of immunity is paid here, in recovery latency and control plane dependence.

\begin{table}[t]
\centering
\caption{Failure recovery: Fat-Tree with ECMP versus \sys.}
\label{tab:failover}
\begin{tabular}{lll}
\toprule
Aspect & Fat-Tree + ECMP & \sys \\
\midrule
Steady state cost & permanent Braess tax (\cref{sec:multipath}) & none \\
Core link fault domain & spread across all pairs & $1/m^2$ of pairs \\
Recovery mechanism & hash recomputation & controller reroute \\
Recovery latency & microseconds & milliseconds \\
Recovery locus & distributed & centralized \\
Path length after failure & unchanged & 2 $\rightarrow$ 4+ hops \\
\bottomrule
\end{tabular}
\end{table}

\subsection{Why the Trade Favors LLM Serving}
\label{sec:pricetrade}

Three properties of LLM workloads make the transient cost cheaper than the permanent tax. First, failures are rare: datacenter network hardware fails at annual rates below 1\%, so \sys runs at its optimum well over 99\% of the time and pays the recovery cost only in the remaining fraction. Fat-Tree pays its Braess tax every day, failure or not. Second, LLM serving stacks already absorb transient network interruptions in upper layers: inference requests retry or migrate across instances, KV cache state can be replicated across Prefill nodes, and training jobs checkpoint periodically~\cite{eisenman2022checknrun,patel2024splitwise}. A millisecond reroute is invisible to mechanisms that already tolerate request level retries. Third, the benefit side of the trade is permanent and user visible: the 40.6\% P99 TTFT reduction of \cref{tab:prod} applies to every request, every day. Rational design prices the two sides by their probability: a certain daily gain against a rare, bounded, absorbed cost.

The deeper statement is about where redundancy should live. Fat-Tree keeps redundancy inside the topology, where it doubles as a Braess risk. \sys moves redundancy outside the topology, into spare NIC ports, precomputed detours, and the resilience of upper layers. \sys does not give up fault tolerance; it relocates fault tolerance to layers that do not tax steady state performance.

\section{Evidence for the Analysis}
\label{sec:evidence}

Two bodies of evidence support the analysis. \cref{sec:evprod} reports production measurements from the GLM-5.1 coding inference cluster. \cref{sec:evsim} uses simulation to check the load spreading theorem and the fragility of static hashing against canonical traffic patterns.

\subsection{Production Measurements}
\label{sec:evprod}

\cref{sec:anomalyinference} describes the deployment: a thousand GPU cluster serving GLM-5.1 coding inference, migrated from ROFT to \sys with no change to GPUs, software, or serving configuration. The measured outcomes appear in \cref{tab:prod}: 33\% lower network cost, 15\% higher GPU inference throughput, and 40.6\% lower P99 TTFT. The cluster has remained stable for several weeks of continuous serving.

Two features of these numbers align with the analysis. First, the gains persist: they are properties of every serving day, not of a measurement window, which matches the claim of \cref{sec:multipath} that the penalty on ROFT is structural rather than stochastic. Second, the tail latency improves more than the mean throughput (40.6\% versus 15\%), which matches the mechanism of \cref{sec:imspreading}: structural congestion is a tail phenomenon, because the overloaded links determine the slowest KV cache transfers, and TTFT waits for the slowest transfer. Finer telemetry, including TTFT distributions and per link utilization before and after migration, is in preparation for a future revision.

\subsection{Simulation: Link Load Imbalance}
\label{sec:evsim}

We check the theorems of \cref{sec:multipath,sec:immunity} on a simulated fabric of $n = 128$ GPUs with $k = 16$ GPUs per switch and $m = 8$ switches per group. ROFT attaches GPU $i$ to Leaf $\lfloor i/k \rfloor$ and hashes each flow to one of $m$ Spine switches, modeling ECMP; we repeat each measurement over 500 random hash seeds. \sys computes core link loads deterministically from \cref{eq:load}. The imbalance metric is the maximum link load divided by the mean link load, so 1.0 denotes perfect balance. We evaluate four traffic patterns: uniform AllReduce, Prefill to Decode transfer with the first 25\% of GPUs as Prefill, the same with 50\%, and 8 random elephant flows.

\begin{figure}[t]
\centering
\includegraphics{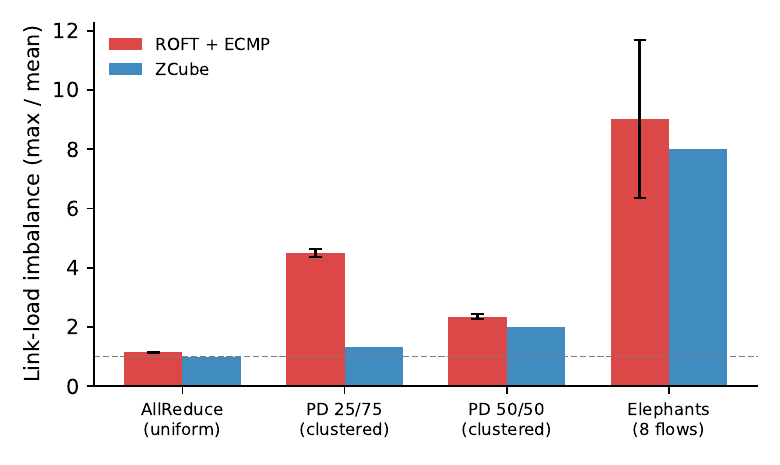}
\caption{Link load imbalance (max over mean) for ROFT with ECMP and \sys under four traffic patterns. Error bars show one standard deviation over 500 ECMP hash seeds. Lower is better; 1.0 is perfect balance.}
\label{fig:imbalance}
\end{figure}

\cref{fig:imbalance} confirms the analysis in three ways. First, under uniform AllReduce traffic both fabrics are nearly balanced, and ECMP still pays a residual imbalance of 1.14 from hash variance while \sys achieves exactly 1.00. Second, under clustered Prefill traffic the fabrics diverge sharply: ROFT reaches an imbalance of 4.50 in the 25/75 case because the contiguous Prefill block sits behind few Leaf switches, while \sys stays at 1.33. The remaining \sys gap consists entirely of legitimately idle links, namely core links toward even groups that contain no Decode nodes for that pattern; restricted to links that carry the pattern's traffic, \sys achieves exactly 1.00 in both Prefill cases, as \cref{sec:imspreading} predicts. Third, under adversarial elephant flows both fabrics degrade to an imbalance near 9, which is the honest boundary of the guarantee: Theorem 3 disperses structured clusters, not arbitrary point flows. The fabrics differ in kind rather than degree here. ECMP draws its imbalance randomly per hash seed, with a standard deviation of 2.67 across seeds, so rehashing permutes the damage without reducing its expectation. \sys computes its assignment deterministically, so an operator can predict and avoid collisions at placement time.

The simulation mirrors the production observation. The pattern that ROFT handles worst, clustered Prefill sources, is precisely the traffic of disaggregated inference, and the pattern on which \sys is exactly uniform on its loaded links is the traffic that the GLM-5.1 cluster serves every day.

\section{Discussion}
\label{sec:discussion}

\noindent\textbf{When does the analysis apply?} The three parts of our analysis identify the regime in which a deterministic fabric beats a multipath one. The traffic must be structured and knowable in advance, so that a fixed route assignment can be good for every pattern it will meet. The routing alternative must be congestion blind or slower than the traffic dynamics, so that multipath cannot realize its theoretical advantage. The traffic volume must concentrate on few large flows, so that hash based spreading fails. LLM training and inference satisfy all three conditions because the model architecture, the parallelism plan, and the serving layout determine the traffic matrix before the first packet flows. General multi-tenant clouds satisfy none of them, and our analysis also explains why the multipath doctrine remains correct there: the assumptions of \cref{sec:mpassumptions} hold, and ECMP works as designed.

\noindent\textbf{Relation to other doctrine violations.} \sys is not the only recent design to question the multipath doctrine. Amazon's RNG replaces the Clos hierarchy with quasi-random graphs and restores path diversity through a distributed routing protocol that finds many edge disjoint paths~\cite{bernardi2026rng}. RNG and \sys attack the same structural problem from opposite sides: RNG keeps many paths and makes routing smarter, while \sys makes the topology smarter and removes route choice. Our analysis predicts that both outperform static ECMP under structured traffic, and that the choice between them hinges on the price of failover that \cref{sec:price} quantifies.

\noindent\textbf{Predictions.} The load spreading theorem is pattern agnostic, so the analysis makes predictions beyond the workloads we measured. Mixture of experts models generate all to all expert routing traffic whose matrix changes across layers and steps; because the guarantee holds for every traffic matrix, the changing matrix should not create structural hotspots on \sys. Colocated training and inference superimposes the two matrices, and the superposition is again a traffic matrix, so the same guarantee applies. Both predictions are testable on the production cluster.

\noindent\textbf{Limitations.} Our analysis has four limits. First, the complete bipartite core requires switch radix to grow with $m$, so very large clusters need hierarchical extensions of the same idea. Second, the $K_{m,m}$ core has substantial wiring complexity, which our deployment handles with automated layout and validation tooling. Third, failure recovery depends on a centralized controller, a single point of operational risk that \cref{tab:failover} states plainly. Fourth, our production evidence comes from one cluster serving one model family, and we attribute the throughput and latency gains to the topology through the zero change migration rather than through a full decomposition of contributing factors.

\section{Related Work}
\label{sec:related}

\noindent\textbf{Greedy routing and the Price of Anarchy.} Braess exhibits the paradox that adding a link can hurt every user~\cite{braess1968}. Roughgarden and Tardos bound the Price of Anarchy of greedy routing at $4/3$ for affine latencies~\cite{roughgarden2002}, and Roughgarden characterizes vulnerable networks through Braess subgraphs~\cite{roughgarden2006}. Routing game theory and datacenter topology design have developed as disjoint literatures, and we are not aware of prior work that applies Braess's paradox to datacenter topology analysis. We apply this theory in reverse: rather than bounding the damage of path multiplicity, we analyze a topology that removes the multiplicity and quantify what it gains and what it pays.

\noindent\textbf{Datacenter topologies.} Fat-Tree and Clos fabrics provision full bisection bandwidth through path multiplicity~\cite{fattree,vl2}, and rail optimized variants tune the hierarchy for GPU collectives~\cite{yan2025atop}. Amazon's RNG replaces the hierarchy with quasi-random graphs and a custom routing protocol~\cite{bernardi2026rng}. \sys removes the Spine layer entirely~\cite{yan2025atop}. Yan et al. present \sys as a system for large model training and evaluate it on training workloads; our work is complementary. We explain why the design works through Braess's paradox, prove the load spreading property that covers arbitrary traffic matrices, and supply the first public production evidence for LLM inference.

\noindent\textbf{Load balancing and adaptive routing.} A long line of work makes multipath routing congestion aware: CONGA balances flows per leaf pair~\cite{conga}, Presto splits flows into cells~\cite{he2015presto}, DRILL routes per packet~\cite{ghorbani2017drill}, packet spraying distributes packets across paths~\cite{packetspraying1}, and MPTCP moves traffic between paths at the transport layer~\cite{mptcp}. These mechanisms attack the same problem as \sys from the routing side. \cref{sec:multipath} gives the precise gap they close: static ECMP never deviates, while these mechanisms deviate with varying latency and overhead. Our analysis suggests that when the traffic matrix is knowable, topology side fixes dominate routing side fixes because they remove the problem rather than react to it.

\noindent\textbf{LLM serving systems.} Disaggregated serving separates Prefill and Decode onto different resources~\cite{patel2024splitwise,zhong2024distserve}, and production stacks schedule KV cache transfers across the cluster~\cite{qin2024mooncake,vllm}. This literature treats the network as given and optimizes placement and scheduling above it. Our results show the two layers interact: the topology determines whether disaggregation traffic congests structurally, and a fabric that matches the traffic's structure improves the serving metrics that this literature optimizes, in our case P99 TTFT by 40.6\%.

\section{Conclusion}
\label{sec:conclusion}

\sys presents the datacenter community with an anomaly: a topology that removes the Spine layer, eliminates path multiplicity, and cuts a third of the network hardware, yet runs LLM training and inference faster. The anomaly traces to a structural root shared with Braess's paradox: congestion-oblivious routing over competing paths. Braess showed that adding paths under this condition can hurt; \sys shows that removing paths under the same condition can help, a connection discovered rather than designed. Multipath fabrics under structured LLM traffic operate in Braess's shadow, where congestion blind hashing is strictly more fragile than the greedy routing that theory bounds. \sys escapes the shadow by construction: its unique path topology is immune to the paradox, and its orthogonal dual partition makes the unique path provably balanced for arbitrary traffic matrices. The price of immunity is real but bounded, and LLM serving stacks already absorb it in upper layers. \sys did not get faster despite having fewer paths. It got faster because of them. For workloads whose traffic derives from model structure, matching topology to traffic is the first order design variable, and path multiplicity is only a means, not the goal.


\bibliographystyle{plain}
\bibliography{refs,nref}

\end{document}